\documentclass[aps,prb,twocolumn,floatfix,footinbib]{revtex4}
\usepackage{epsfig}
\usepackage{amsfonts}
\usepackage{amsmath}
\usepackage{amssymb}
\usepackage{graphicx}%
\begin{document}
\title{Analytical Model of Spin-Polarized Semiconductor Lasers}
\author{Christian G\o thgen$,^{1}$ Rafal Oszwa\l dowski,$^{1,2}$ 
Athos Petrou,$^{1}$ and Igor \v{Z}uti\'{c}$^{1}$}
\affiliation{
$^{1}$ Department of Physics, State University of New York 
at Buffalo, NY, 14260, USA \\
$^{2}$ Instytut Fizyki, Uniwersytet Miko{\l }aja Kopernika, Grudzi\c{a}dzka
5/7, 87-100, Toru\'n, Poland}

\begin{abstract}
We formulate an analytical model for vertical-cavity surface-emitting lasers
(VCSELs) with injection (pump) of spin-polarized electrons. Our results for
two different modes of carrier recombination allow for a systematic analysis
of the operational regimes of the spin-VCSELs. We demonstrate that threshold
reduction by electrically-pumped spin-polarized carriers can be larger than
previously assumed possible. Near the threshold, such VCSELs can act as
effective non-linear filters of circularly-polarized light, owing to their 
spin-dependent gain.
\end{abstract}
\maketitle

Spin-dependent properties have been successfully used in metallic magnetic 
multilayers for a variety of device applications exploiting magnetoresistive 
effects.~\cite{Zutic2004:RMP} Unfortunately, much less is known about practical 
paths to implement spin-controlled devices that would go beyond 
magnetoresistance.~\cite{Zutic2004:RMP,Fabian2007:APS} One such 
encouraging development is demonstrated with spin-polarized vertical cavity 
surface-emitting lasers (spin-VCSELs).~\cite{Rudolph2003:APL,Holub2007:PRL} 
Spin-polarized carriers created by circularly-polarized photo-excitation or 
electrical injection can 
enhance the performance of spin-VCSELs as compared 
to their conventional (spin-unpolarized) counterparts. 
This work~\cite{Rudolph2003:APL,Holub2007:PRL} has   
demonstrated threshold current reduction and independent modulation of 
optical polarization and intensity. While numerical results using the rate 
equation (RE) description of spin-VCSELs were already 
presented,~\cite{Rudolph2003:APL,Holub2007:PRL} a large number 
of materials parameters makes it difficult to systematically elucidate how 
the spin-dependent effects modify the device operation. Consequently, even 
the widely accepted theoretical limit of maximum 50\% threshold current 
reduction needs to be re-examined.

For clarity  
of our approach we formulate a simple RE~\cite{Dery2004:IEEEJQE}
model of spin-VCSELs which, while similar to prior numerical 
work,~\cite{Rudolph2003:APL,Holub2007:PRL} can be solved analytically. 
We consider a quantum well (QW) 
as the active region in the VCSEL. 
Spin-resolved electron and hole densities are $n_\pm$, $p_\pm$,
where $+(-)$ denotes the spin up (down) component; the total carrier 
densities are $n=n_++n_-$, $p=p_++p_-$. Analogously, for photon 
density we write $S=S^++S^-$, where $+(-)$ is the right (left) 
circularly-polarized component.~\cite{convention}  
Electrically or optically injected/pumped spin-polarized electrons into 
the QW can be represented by  
current density $J=J_++J_-$ (normalized 
to unit charge and an effective volume in the QW region~\cite{Holub2007:PRL}).
Typically, the spin relaxation time of holes is much shorter than for 
electrons,\cite{Zutic2004:RMP} $\tau_s^p  \ll \tau_s^n$, 
implying that the holes can be considered unpolarized with $p_\pm=p/2$.

In conventional VCSELs, the optical gain term, describing stimulated 
emission, 
can be simply modeled as
$g(n,S)=g_0(n-n_{\mathrm{tran}})/(1+\epsilon S)$, where $g_0$ is the 
density-independent coefficient,\cite{Holub2007:PRL} 
$n_{\mathrm{tran}}$ is the transparency density, 
and $\epsilon$ is the gain compression factor. We generalize this relation 
in the spin-polarized case as $g(n,S) \rightarrow g_\pm(n_\pm,p_\pm,S^\pm)=%
g_0(n_\pm+p_\pm-n_{\mathrm{tran}})/(1+\epsilon S)$.
This form differs from the previously employed gain 
expressions,~\cite{Rudolph2003:APL,Holub2007:PRL} as it explicitly
contains the hole density, but it coincides with the rigorously derived gain 
expression from semiconductor Bloch equations.~\cite{Chow:1999}

The charge neutrality $p_\pm=p/2=n/2$, allows us to recover the 
spin-unpolarized limit for the gain expression, as well as to decouple the 
REs for electrons from 
those for holes. The spin-polarized REs for electrons thus become
\begin{eqnarray}
dn_\pm/dt=J_\pm-g_\pm(n_\pm,S^\mp)S^\mp
  -\left(n_\pm-n_\mp\right) /\tau_s^n
  -R_{\mathrm{sp}}^\pm, \label{Eq.ab} \\
dS^\pm/dt=\Gamma g_\mp(n_\mp,S^\pm)S^\pm 
-S^\pm/\tau_{\mathrm{ph}}+\beta\Gamma R_{\mathrm{sp}}^\mp, 
\quad \quad \quad \quad \label{Eq.ac}
\end {eqnarray}
where $\tau_{\mathrm{ph}}$ is the photon lifetime, $\Gamma$ is the optical
confinement coefficient, $\beta$ is the spontaneous-emission coupling 
coefficient, and $R_{\mathrm{sp}}^{\pm}$ 
is the radiative spontaneous recombination rate.
For the spin-unpolarized case, $R_{\mathrm{sp}}^\pm$ 
is typically assumed to be quadratic in carrier densities,~\cite{valid} 
$R_{\mathrm{sp}}=B np/2$, 
where $B$ is a temperature-dependent constant.~\cite{Chow:1999} 
Another simple 
form, $R_{\mathrm{sp}}=n/\tau_{\mathrm{r}}$,
where  $\tau_{\mathrm{r}}$ is the recombination time,  
is a good approximation at high $n$,~\cite{Bourdon2002:JAP}
but was not considered in the prior work on 
spin-VCSELs.~\cite{Rudolph2003:APL,Holub2007:PRL}
For spin-polarized electrons these forms are generalized 
as~\cite{Zutic2006:PRL}
$R_{\mathrm{sp}}^\pm=B n_\pm p_\pm=B n_\pm n/2$ 
and $R_{\mathrm{sp}}^{\pm}=n_{\pm}/\tau_{\mathrm{r}}$. We 
refer to them as quadratic and linear recombination (QR and LR),
which can be viewed as the 
opposite limits of a general expression for $R_{\mathrm{sp}}^{\pm}\left(
n_{\pm},p_{\pm}\right)$.~\cite{Auger}

To describe the solutions of Eqs.~(\ref{Eq.ab},\ref{Eq.ac}), which are the basis 
of our theoretical approach, we introduce polarizations of injected electron
current $P_J=(J_+ - J_-)/J$ as well as of electron and photon densities 
$P_n=(n_+-n_-)/n$, $P_S=(S^+-S^-)/S$. Electrical injection in QWs, 
using Fe or FeCo, allows 
for~\cite{Zutic2004:RMP,Hanbicki2002:APL} $|P_n|\sim 0.3-0.7$ with
similar values for $|P_J|$, while $|P_n|\rightarrow 1$ is
attainable optically at room temperature.~\cite{Zutic2004:RMP,negative} 
In the  unpolarized limit
$J_+=J_-$, $n_+=n_-$, $S^+=S^-$ or $P_J=P_n=P_S\equiv0$. As in the 
recent experiments,~\cite{Holub2007:PRL} we consider a continuous wave 
operation of VCSEL and look for steady-state solutions of  
Eqs.~(\ref{Eq.ab},\ref{Eq.ac}). Guided by the experimental 
range~\cite{Rudolph2003:APL,Holub2007:PRL} of $\beta\sim10^{-5} -10^{-3}$,
we mostly focus on the limit $\beta = 0$, for which all the 
operating regimes of spin-VCSELs can be simply described, and, additionally,
consider $\epsilon = 0$, relevant for moderate pumping intensities. 
\begin{figure}[tbh]
\includegraphics[ scale=0.7]{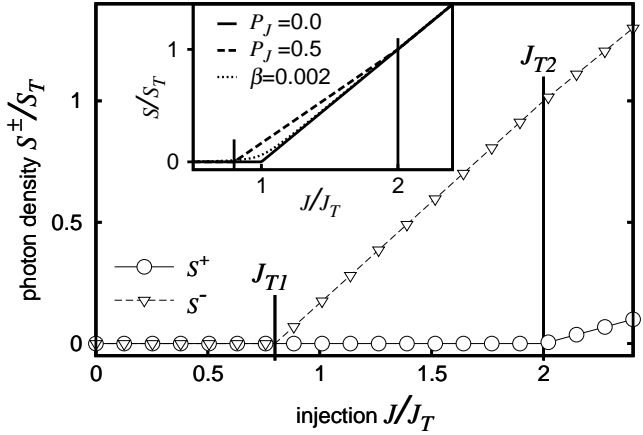} 
\caption{Photon densities of the left- ($S^-$) and right- ($S^+$) circularly 
polarized light as a function of electron current $J$ with polarization 
$P_{J}=0.5$, infinite electron spin relaxation time and linear recombination. 
$J$ is normalized to the unpolarized threshold current $J_T$ and $S^\pm$
to $S_T=J_{T}\Gamma\tau_{\mathrm{ph}}$.
The vertical lines indicate $J_{T1}$ and $J_{T2}$ thresholds.
Inset: total photon density ($S=S^++S^-$) is shown
for spin-unpolarized laser ($P_J=0)$ and two spontaneous-emission
coupling coefficients ($\beta=0, 0.002$), as well as a spin laser
with $P_J=0.5$,  $\beta=0$.}
\label{Fig.reg_def}
\end{figure}

We first show analytical results for \emph{unpolarized} ($P_{J}=0$) VCSEL and
LR in the inset of Fig.~\ref{Fig.reg_def}.
Injection current density is normalized to the
unpolarized threshold value, $J_{T}=N_{T}/\tau_{\mathrm{r}}$, with $N_{T}$ 
denoting the total electron
density at (and above) the threshold, $N_{T}=n\left( J\geq J_{T}\right)  
=\left(  \Gamma g\tau_{\mathrm{ph}}\right)^{-1}+n_{\mathrm{tran}}$,
while photon density is normalized to  $S_T=J_T \Gamma \tau_{ph}$.
A small difference in $S(J)$ between the vanishing (solid line)
and finite $\beta$ (dotted line, $\beta=0.002$ overestimates the 
experimental values~\cite{Rudolph2003:APL,Holub2007:PRL}) shows 
the accuracy of $\beta=0$ approximation.
The unpolarized case has two regimes: for $J<J_T$  the device behaves as a 
light-emitting diode (LED), with negligible stimulated emission; 
for $J>J_T$, it is a fully lasing VCSEL. 

With finite $P_J$ and $\tau_s^n \rightarrow \infty$, we reveal a 
more complicated 
behavior further 
explored  
in Fig.~\ref{Fig.reg_def}, main panel. The two
threshold currents, $J_{T1}$ and $J_{T2}$ 
($J_{T1}\leq J_T\leq J_{T2}$, the equalities hold only when $P_J=0$),
delimit three regimes of a spin-VCSEL.   
(i) For $J<J_{T1}$ it operates 
as a spin-LED.~\cite{Zutic2004:RMP,Hanbicki2002:APL}
(ii) For $J_{T1}\leq J_T\leq J_{T2}$, there is mixed operation: lasing
only with left-circularly polarized light $S^-$ 
(we assume~\cite{negative} $J_+ >J_-$), 
which can be deduced from
the spin-dependent gain term in Eqs.~(\ref{Eq.ab},\ref{Eq.ac}), while
$S^+$ is still in a spin-LED regime ($S^+ \rightarrow 0$ for 
$\beta \rightarrow 0$). (iii)   
For $J \ge J_{T2}$, it is fully lasing with both $S^\pm >0$.

RE description of spin-unpolarized lasers reveals that the carrier densities are
clamped above $J_T$. We have found a related effect in the spin-polarized
case. For $J>J_{T1}$, Eq.~(\ref{Eq.ac}) for $S^-$ can be divided by
$S^-$,  
showing that the quantity
$n_++p_+=n_++n/2$ 
(we recall the charge neutrality 
condition and $\tau_s^p \rightarrow 0$)  
will be clamped at $N_T$. 
If $J_{T1}<J<J_{T2}$ then neither $n_+$ nor $n_-$ are separately clamped. 
If $J>J_{T2}$, then $n_-+n/2=N_T$ must hold in
addition to the previous condition $n_++n/2=N_T$. These two conditions yield 
$n_\pm=N_T/2$, independent of $P_J$. Thus, above $J_{T2}$, the sum of 
Eqs.~(\ref{Eq.ac}) for $S^-$ and $S^+$ reduces to the usual 
unpolarized equation, and $S$ is independent of $P_J$ (inset of 
Fig.~\ref{Fig.reg_def}). However, if $P_{J}\neq0$ then we still find
$S^-\neq S^+$ (Fig.~\ref{Fig.reg_def}, $J>J_T$).

\begin{figure}[tbh]
\includegraphics[scale=0.7]{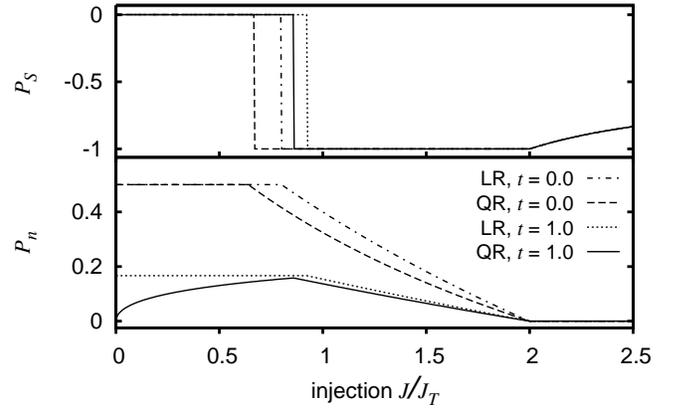} 
\caption{Electron ($P_n$) and
photon ($P_S$) polarizations as a function of injection current $J$, 
with polarization 
$P_J=0.5$ and different ratios of recombination and electron spin relaxation time, $t$. 
Linear (LR) and quadratic (QR) recombination  
are shown.
$J$ is normalized to the unpolarized threshold  
$J_T$.
}
\label{Fig.polars}
\end{figure}

Most of the results discussed above do not change qualitatively for a finite 
$\tau_s^n$  or QR. The general features, such as the existence of
three regimes of operation, remain the same. However, 
the photon and carrier densities (for a given $J$) as well as the threshold
$J_{T1}$ depend quantitatively 
on the spin-flip rate and the recombination form. 
We investigate this dependence in Fig.~\ref{Fig.polars} which shows 
the evolution of photon and carrier polarizations with the injection current.
We consider both LR and QR forms and express our results
using the ratio of the radiative recombination and spin relaxation times:
$t=\tau_{\mathrm{r}}/\tau_s^n$. For QR the unpolarized threshold is
$J_{T}=B N_T^2/2$, and we define 
$\tau_{\mathrm{r}}=N_T/J_T$ by analogy with the LR case.  

When $0<J<J_{T1}$, then $P_n(J)$  depends on the recombination 
form; $P_n$ is constant for LR, while for QR it grows 
monotonically.~\cite{Meier:1984} 
Only for a very long spin relaxation time, 
$t\rightarrow0$, $P_n(J<J_{T1})  \rightarrow P_J$. 
At $\beta =0$ 
there is no stimulated emission, so that $P_S\equiv0$.
When $J_{T1}<J<J_{T2}$, $P_S=-1$ for any  $t$, recall
Fig.~\ref{Fig.reg_def}.  
With only a partially polarized electron current, the spin-VCSEL emits 
fully circularly-polarized light,  analogous to the spin-filtering effect
in magnetic materials. In the same regime, $P_{n}$ decreases with $J$ 
because, while
$n_-$ grows, $n_+$ must drop in order to maintain 
$n_++n/2=3/2n_++1/2n_-=N_{T}.$

For $P_{J} \neq 0$, the $J_{T2}-J_{T1}$ interval is widest for 
$\tau _s^n \rightarrow \infty$. 
When $\tau_s^n$ decreases then the lower threshold $J_{T1}$ grows
towards its unpolarized counterpart $J_{T}$, and $J_{T2}-J_{T1}$ contracts,
Fig.~\ref{Fig.polars}. 
We find $J_{T2}=J_{T}/(1-P_{J})$, independent of $\tau_s^n$, 
this should hold up to very small values of $\tau_s^n$
($t\sim10^2$). In the extreme case of even smaller $\tau_s^n$, 
for  $J_{T1}<J<J_{T2}$
the majority and minority spin densities become almost equal; 
$n_+\gtrsim N_T$ and $n_-\lesssim N_T$, so that $\beta \sim 10^{-4}$
can effectively drive the device to the fully lasing regime. 
For typical values~\cite{Rudolph2003:APL,Holub2007:PRL} 
of $t\sim 5$, setting $\beta=0$ is an accurate assumption. 
At $J_{T2}$, we recall $n_+=n_-(=N_T/2)$ so that the
spin-flip term in Eq.~(\ref{Eq.ab}) for $n_-$ is zero. The equation reduces to
$J_-\equiv1/2(1-P_J)J=R_{\mathrm{sp}}^-$, explaining why
$J_{T2}$ depends only on $P_J,$ but not on $t$.  For the same reason,
$P_S$ and $P_n$ are independent of $t$ when $J\geq
J_{T2}$. The quantity $P_S$ increases ($\left\vert P_{S}\right\vert $
decreases) with $J$ 
as $P_S= -P_J J/(J-J_T)$
to the asymptotic value $P_S=-P_J$. Thus $P_J$ can be inferred from $P_S$
only at sufficiently high injection.
In all three regimes, defined by Fig.~\ref{Fig.reg_def}, we find that $0\leq
P_n<P_J$, (only for $\tau_s^n\rightarrow\infty$ and $J<J_{T1}$,
$P_n\rightarrow P_J$). 
For $J>J_{T2}$ we find $P_n=0$.
Therefore $P_n$ is not enhanced in the spin-VCSEL's 
active region.

\begin{figure}[tbh]
\includegraphics[scale=0.7]{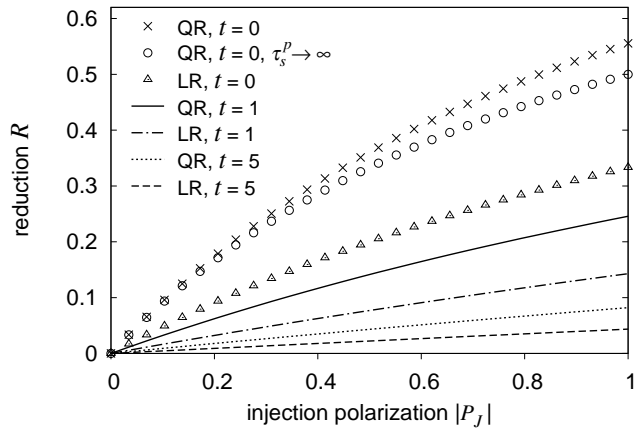}
\caption{Threshold current reduction $R$ as a function of injection 
polarization $P_J$. Shown are both linear and quadratic recombination
(LR and QR) for various $t$, the ratio of recombination time and
electron spin relaxation time. The hole spin relaxation time $\tau _s^p=0$, 
except for the dashed line and what was previously assumed ``ideal'' case, 
in which $\tau_s^p\rightarrow\infty$.}
\label{Fig.em_enh}
\end{figure}

Experiments have demonstrated~\cite{Rudolph2003:APL,Holub2007:PRL}  
that injecting spin-polarized carriers reduces the threshold current 
in a VCSEL, i.e.,  $J_{T1}(P_J\neq0)<J_T$.  We quantify this 
threshold-current reduction~\cite{Rudolph2003:APL,Holub2007:PRL}  
with $R=(J_T-J_{T1})/J_T$, obtained analytically from REs
\begin{align}
\text{LR: } & R=\left\vert P_{J}\right\vert /(2+\left\vert
P_{J}\right\vert +4t),\label{Eq.af}\\
\text{QR: } & R=1-2/\left(2+\left\vert P_{J}\right\vert \right)
^{2}\times\label{Eq.ag}\\
& \left[  1+2\left\vert P_{J}\right\vert t+4t^{2}+\left(  1-2t\right)
\sqrt{\left(  1+2t\right)  ^{2}+4\left\vert P_{J}\right\vert t}\right],
\nonumber
\end{align}
and we find $R>0$, for any $P_J\neq0$, $\tau_s^n>0$ and both LR and QR.
As expected, $R\rightarrow0$ for $\tau_s^n\rightarrow0$ ($t\rightarrow\infty$), 
independent of the recombination form, since we have always assumed $\tau_s^p=0$.
We illustrate further $R(P_J)$ for both LR and QR 
in Fig.~\ref{Fig.em_enh}. 
We choose $t=1,5$, close to the typical values 
$\tau_{\mathrm{r}}\sim1$ ns and $\tau_{\mathrm{s}}\sim100$ ps for QWs
at room temperature.~\cite{Rudolph2003:APL} 
From Fig.~\ref{Fig.em_enh} or
Eqs.~(\ref{Eq.af},\ref{Eq.ag}) one can see that $R(P_J,t)$ for LR is always
smaller than for QR. 
For $\left\vert P_J\right\vert =1$ and $\tau_s^n \rightarrow \infty$
(but still $\tau_s^p=0$)  we obtain the maximum
current reductions
\begin{equation}
\text{LR} \text{: }R=1/3, \quad 
\text{QR} \text{: }R=5/9\text{.} \label{Eq.ae}
\end{equation}
Remarkably, for QR the maximum $R$ is larger than previously assumed 
possible $R=1/2,$~\cite{Rudolph2003:APL,Holub2007:PRL} even though holes 
are completely unpolarized ($\tau_s^p=0$). This can be explained as follows.
First we calculate $n$ from Eq.~(\ref{Eq.ac}) for $S^-$.
The threshold density is reached when $n_++n/2=N_T$, which gives
$n=(2/3)N_T$ for $P_J=1,$ i.e., for $n_+=n$. Thus the
threshold \emph{electron density} is reduced by only $1/3$. Assuming $S^-=0$
at the threshold, the threshold current from Eq.~(\ref{Eq.ab}) is given
by $J_+=B n_+n/2$. If $P_J=1$ then 
$J=J_+=B n^2/2=(2/3)^{2}J_T$, 
which yields the $5/9$
\emph{current} reduction, a direct consequence of the quadratic
dependence of recombination on $n$ (note that for LR both the
$N_T$ and $J_T$ are reduced only by $1/3$). 

For comparison, we  
calculate  
$R$ for QR in a priori ``ideal'' case,
where $n_{\pm}=p_{\pm}$ for any injection $J$ and
\emph{both} $\tau_s^n, \tau_s^p \rightarrow \infty$. Therefore the gain and
QR terms in Eq.~(\ref{Eq.ab}) for $n_+$ ($n_-$) and in
Eq.~(\ref{Eq.ac}) for $S^-$ ($S^+$) must be modified by replacing
$n/2$ with $n_+$ ($n_-$). Surprisingly, we obtain $R\left(  \left\vert
P_J\right\vert =1\right)  =1/2$, a smaller reduction than in
the  case of unpolarized holes. A simple calculation 
yields $n=N_T/2$, i.e., the reduction of the threshold
\emph{electron density} is larger than for $\tau_s^p=0$. 
However, the interband recombination $R_{\mathrm{sp}}^+$ is more efficient,
because none of the holes undergo spin-flip. Thus we obtain
$J_+=J_{T1}=B n_+^2=B N_T^2/4=J_T/2$. 
For $\tau_s^p=0$ and $P_J=1$, half of the holes are inaccessible
for recombination with the fully spin-polarized electrons. This lowers the
injection necessary to overcome the recombination losses and the interplay
of stimulated and spontaneous recombination leads to a smaller
$R$ in the ``ideal'' case.

In this work we show that even a simple rate equation model can reveal
surprising trends for operation of spin-VCSELs. The maximum threshold
current reduction is not achieved for infinite electron and hole spin 
relaxation times, but rather when the hole spin lifetime vanishes. 
The corresponding threshold reduction exceeds the previously considered 
theoretical limit. We expect that 
the transparency of our analytical approach will help to 
elucidate the operation of spin-controlled 
lasers, which continue to be actively studied.~\cite{Hovel2008:APL} 

This work was supported by the U.S. ONR and NSF-ECCS Career. We thank
M. Holub, M. Oestreich, and G. Strasser for stimulating discussions.


\begin{thebibliography}{99}
\bibitem{Zutic2004:RMP}
I. \v{Z}uti\'c {\em et al.}, Rev. Mod. Phys. {\bf 76}, 323 (2004).

\bibitem{Fabian2007:APS}
J. Fabian {\em et al.}, Acta Phys. Slov. {\bf 57}, 565 (2007).

\bibitem{Rudolph2003:APL}
J. Rudolph {\em et al.}, Appl. Phys. Lett. {\bf 82}, 4516 (2003);
Appl. Phys. Lett. {\bf 87}, 241117 (2005).

\bibitem{Holub2007:PRL}
M. Holub {\em et al.}, Phys. Rev. Lett. {\bf 98}, 146603 (2007);
J. Phys. D: Appl. Phys. {\bf 40}, R179 (2007).

\bibitem{Dery2004:IEEEJQE}
Generally, REs can provide a versatile description for a variety of
lasers, for example,
H. Dery and G. Eisenstein, IEEE J. Quant. Electron. {\bf 40}, 1398 (2004).

\bibitem{convention}
We use ``the angular momentum convention,'' described in B.~T. Jonker 
{\it et al.,} J. Magn. Magn. Matter {\bf 277}, 24 (2004). 

\bibitem{Chow:1999}
W.~W. Chow and S.~W. Koch, Semiconductor-Laser Fundamentals: Physics of the 
Gain Materials (Springer, 1999). 

\bibitem{valid}
This form is strictly valid if both quasi-Fermi levels are deep in the gap, 
i.e., below the inversion condition. 

\bibitem{Bourdon2002:JAP}
G. Bourdon {\em et al.}, J. Appl. Phys. {\bf 92}, 6595 (2002).

\bibitem{Zutic2006:PRL}
I. \v{Z}uti\'c {\em et al.}, Phys. Rev. Lett. {\bf 97}, 026602 (2006);
equilibrium densities are negligibly small for VCSEL operation.

\bibitem{Auger}
For the injection levels considered, Auger recombination is not significant.

\bibitem{Hanbicki2002:APL}
A.~T. Hanbicki {\it et al.}, Appl. Phys. Lett. {\bf 80}, 1240 (2002);
T.~J. Zega {\it et al.}, Phys. Rev. Lett. {\bf 96}, 196101 (2006); 
G. Salis {\it et al.}, Appl. Phys. Lett. {\bf 87}, 262503 (2005). 

\bibitem{negative}
We consider $P_J\ge0$ while the results for $P_J<0$ can deduced easily.
Other approaches were considered for $P_J\equiv0$.
M. San Miguel {\em et al.}, Phys. Rev. A {\bf 52}, 1728 (1995).

\bibitem{Meier:1984}
Similar behavior is known for bulk-like semiconductors. 
{\it Optical Orientation,} edited by F. Meier and B.~P. Zakharchenya
(North-Holland, New York, 1984);   
I. \v{Z}uti\'c {\em et al.}, Phys. Rev. B {\bf 64}, 121201(R) (2004).

\bibitem{Hovel2008:APL}
S. H\"{o}vel {\it et al.}, Appl. Phys. Lett. {\bf 92}, 041118 (2008).
  
\end{thebibliography}
\end{document}